*Article*

# Inclusive STEAM Education: A Framework for Teaching Coding and Robotics to Students with Visually Impairment Using Advanced Computer Vision

**Mahmoud Hamash [1*], MD RAQIB KHAN [2] and Peter Tiernan [1,]**

1. Affiliation 1; School of STEM, Innovation & Global Studies, Institute of Education, Dublin City University, mahmoud.hamash2@mail.dcu.ie
2. Affiliation 2; School of Computer Science and Statistics Trinity College Dublin, khanmd@tcd.ie
* Correspondence mahmoud.hamash2@mail.dcu.ie

**Abstract:** STEAM education integrates Science, Technology, Engineering, Arts, and Mathematics fostering creativity and problem-solving, yet students with Visual Impairments (VI) face challenges in programming and robotics, particularly in tracking a robot's movements and spatial awareness. This paper proposes a framework that integrates pre-constructed robots and algorithms, such as maze-solving techniques, into an accessible learning environment. Using CLIP (Contrastive Language-Image Pre-training), a global camera captures maze layouts, converting images into textual descriptions that generate spatial audio prompts in an Audio Virtual Reality (AVR) system. Students provide verbal commands, which are processed and refined through CLIP. Additionally, robot-mounted stereo cameras supply real-time data, processed via SLAM (Simultaneous Localization and Mapping) for continuous feedback. This framework enables VI students to develop coding skills and engage in complex problem-solving tasks. Beyond maze-solving, it offers broader applications of computer vision in special education, advancing accessibility and learning in STEAM disciplines.

## 1. Introduction

The integration of Science, Technology, Engineering, Arts, and Mathematics (STEAM) education has gained prominence as a transformative educational approach that nurtures creativity, critical thinking, and interdisciplinary problem-solving skills among students [1,2]. By bridging traditionally siloed disciplines, STEAM equips learners with the tools to navigate and address real-world challenges. However, despite its transformative potential, STEAM education is not universally accessible. Students with Visual Impairments (VI), for instance, face significant barriers to fully engaging with its hands-on, technology-driven methodologies [3-5]. This lack of access not only limits their educational opportunities but also hinders the inclusion of their unique perspectives in STEAM fields.

Programming and educational robotics, as integral components of STEAM education, offer significant opportunities for students to develop computational thinking, problem-solving, and collaboration skills [6,7]. For students with VI, robotics has the potential to serve as a tangible and adaptive tool to explore abstract concepts through real-world applications [8-9, 4]. However, teaching programming and robotics to students with VI presents unique challenges, particularly in enabling real-time awareness of a robot's movements and surroundings, critical aspects for debugging code and achieving successful outcomes [10,11]. Without accessible methods to interact with the robot's environment, these students often remain excluded from the full benefits of robotics education. The challenges faced by students with VI in STEAM education are deeply intertwined with the limitations of existing methodologies and technologies. Traditional approaches often rely on visual interfaces and feedback, rendering them inaccessible to students with VI [12,13]. To address this gap, emerging technologies such as computer vision, natural language processing, and assistive systems must be leveraged to create innovative, inclusive solutions.

Technological advancements offer innovative opportunities to make education more inclusive, engaging, and accommodating [14]. Assistive technologies and inclusive educational tools have proven effective in facilitating STEAM learning for students with disabilities, including those with VI, by leveraging affordable and accessible resources [15]. Among these advancements, Computer Vision (CV) a field of artificial intelligence that enables machines to interpret and process visual data from the world—and image understanding, which focuses on deriving meaningful insights from images through analysis and pattern recognition, have emerged as pivotal technologies [16]. These technologies are widely implemented in assistive tools to support navigation, obstacle detection, and object localisation for individuals with VI [17]. Furthermore, the increasing ubiquity of computing power has bolstered the potential for deploying such innovative solutions at scale [18], breaking down barriers and enhancing educational opportunities for students with VI.

Building on prior research that identified the urgent need for accessible tools in STEAM education, this study seeks to address the challenges students with VI encounter in programming and robotics. Our earlier findings underscored the remarkable problem-solving abilities and creativity of these students, despite being hindered by a lack of accessible technologies tailored to their needs [19, 4]. This work aims to develop a framework that addresses these technical barriers while empowering students with VI to fully participate in STEAM activities, particularly in programming and robotics.

This work's significance lies in its potential to bridge the gap between the unique needs of students with VI and the practical application of programming educational robots. By addressing the barriers identified in prior research, the proposed framework aims to create an inclusive, interactive educational environment that empowers these students and expands their opportunities in STEAM fields. The framework integrates cutting-edge technologies to reimagine STEAM education in a way that is equitable and transformative for all learners.

In summary, this paper emphasises the critical need for inclusive STEAM education and examines the challenges and opportunities in teaching programming and robotics to students with VI. It builds on prior research to lay the groundwork for a transformative framework that leverages technology to create equitable learning experiences. The subsequent sections explore the theoretical underpinnings, methodologies, and solutions driving this framework, offering insights into how STEAM education can be reimagined to benefit diverse learners.

## 2. Literature Review

STEAM education, which incorporates Science, Technology, Engineering, Arts, and Mathematics, is widely regarded as a critical framework for equipping students with 21st-century skills such as critical thinking, problem-solving, creativity, and collaboration [1]. By focusing on interdisciplinary learning, STEAM education prepares students to address complex real-world problems through innovative and holistic approaches. It also nurtures technological proficiency, which is essential in a rapidly evolving digital landscape [2]. However, while the adoption of STEAM education has grown substantially, its implementation often remains inaccessible for students with disabilities, particularly those with VI [11].

For students with VI, engaging in STEAM education presents unique challenges and opportunities. Many concepts in STEAM, such as visual programming, spatial reasoning, and laboratory experimentation, rely heavily on visual and spatial inputs, creating significant barriers for learners with VI [20]. Despite these challenges, STEAM education can provide valuable opportunities for students with VI to develop critical technological and practical skills. As noted by [21], early exposure to STEAM education can significantly influence the attitudes and aspirations of students who are VI, encouraging them to pursue their passions and ambitions. Furthermore, [5] emphasises that ensuring students with VI can actively participate in STEAM education is not only a matter of equity but also an opportunity to leverage their unique sensory and cognitive strengths to approach problem-solving from novel perspectives.

A key challenge in making STEAM education accessible for students with VI lies in the inherent visual nature of many educational tools and practices. Programming and robotics, which form essential components of STEAM curricula, often involve visual coding environments and visual feedback systems. Traditional programming platforms are rarely designed with accessibility in mind, creating barriers for students who rely on non-visual modes of interaction [19, 22]. Another key challenge in STEAM education for students with VI is providing real-time spatial awareness in laboratory and robotics activities [19, 23]. Traditional methods, which rely on visual cues for feedback and navigation, are often inaccessible to students with VI. Our previous research [4] found that students with visual impairments (VI) lose direct connection and feedback when deploying their programmed robots onto a track. For example, when programming a robot to navigate a maze, students with VI rely on their initial observations, such as haptic feedback or input from teachers and sighted peers. However, once the robot is deployed, they no longer receive direct feedback from it, making it difficult to develop spatial awareness of the robot's position in relation to the maze components. To address this challenge, educators often step in by providing real-time observations or encouraging sighted peers to guide students with VI through the process.

However, recent advancements in accessible programming platforms, such as Quorum, Blockly, and MakeCode demonstrate how auditory feedback, keyboard navigation, screen readers, and tactile interfaces can be utilised to make programming education more inclusive [24]. Similarly, tactile robotics kits that incorporate audio feedback have been shown to enable students with VI to actively engage in robotics education, enhancing their learning outcomes and fostering a deeper understanding of STEAM concepts. For instance, Hamash and Mohamed [4] and also, Ludi & Jordan [25] demonstrated that students with visual impairment could successfully build and program robots when provided with adapted materials and guided instruction. And as stated by the researchers this hands-on experience not only boosts their confidence but also cultivates essential skills in technology and engineering.

Furthermore, recent technological developments, such as the integration of computer vision (CV) systems with multimodal feedback mechanisms, offer promising solutions [26]. These innovations have the potential to significantly enhance accessibility in STEAM education by creating interactive and immersive learning environments tailored to the needs of students with VI. Computer vision (CV) technology presents a promising solution to overcome these challenges and create inclusive learning environments in STEAM education. Computer vision (CV) is a multidisciplinary field that enables machines to interpret and understand visual information, mimicking human visual perception [27]. By combining techniques from artificial intelligence (AI), machine learning, image processing, and pattern recognition, computer vision systems analyze and extract meaningful data from images, videos, and other visual inputs [28, 29]. The ultimate goal of the field is to automate tasks that traditionally require human visual cognition, such as object detection, facial recognition, and scene reconstruction [30]. Over the past few decades, computer vision has evolved from simple image-processing tasks to complex systems capable of real-time analysis and decision-making, driven by advancements in algorithms, computational power, and the availability of large-scale datasets [31].

CV systems, which enable computers to interpret and process visual information from the environment, can be adapted to translate visual data into accessible formats for students with VI. For instance, CV-powered tools can provide real-time auditory or haptic feedback, allowing students to engage with spatial and visual content in non-visual ways. For example, Wang et al. [32] illustrated how CV technologies can capture spatial and environmental data and translate it into auditory or tactile feedback, enabling students with VI to navigate and interact with physical spaces more effectively. Similarly, CV has been applied in various innovative ways to support individuals with VI, such as object recognition, text-to-speech systems, and assistive navigation tools.

One notable application is the use of CV for object recognition and description. Tapu et al. [33] developed a wearable assistive device that uses CV to identify objects in the user's surroundings and provide real-time audio descriptions, helping individuals with VI recognise everyday items and navigate unfamiliar environments. Additionally, CV has been integrated into navigation systems, such as the work by Brock and Kristensson [34], who designed a CV-based system that provides auditory guidance to help users with VI navigate indoor and outdoor spaces safely. An additional example is the application of computer vision (CV) in two innovative travel aids designed for blind pedestrians [35]. In the first application, CV is utilised to generate acoustic signals that assist users in navigating around obstacles. However, these audio signals proved to be confusing for users, resulting in increased cognitive load. In contrast, the second application featured a customised guidance cane equipped with a servo steering mechanism and sensors for obstacle detection. This approach was more user-friendly; however, it faced limitations, such as the inability to detect stairs. Additionally, CV has been integrated into navigation systems, such as the work by [34], who designed a CV-based system that provides auditory guidance to help users with VI navigate indoor and outdoor spaces safely.

Finally, CV has been used to enhance educational experiences for students with VI, particularly in STEM education, programming, and robotics. [36] developed a CV-based tool that converts graphical and mathematical content into tactile and auditory formats, making STEM education more accessible. Similarly, [37] created a CV-driven programming environment that uses auditory feedback to help students with VI write and debug code, enabling them to participate in computer science education. In robotics, [10] demonstrated how CV can be integrated into robotics kits to provide real-time feedback through sound and vibration, allowing students with VI to engage in hands-on robotics activities. Furthermore, [38] explored the use of CV to create accessible interfaces for 3D modelling and design, empowering students with VI to participate in engineering and design projects.

Building on the role of CV in enhancing educational experiences for students with VI, particularly in STEM, programming, and robotics, Simultaneous Localization and Mapping (SLAM) emerges as another critical technology that complements these advancements. SLAM enables devices to map unknown environments while simultaneously tracking their position in real-time, making it invaluable for applications like autonomous navigation. Recent improvements in SLAM algorithms, such as Extended Kalman Filter SLAM (EKF-SLAM) [55] and ORB-SLAM [56], have significantly enhanced localization accuracy and robustness, allowing their integration into educational tools that provide real-time spatial awareness for students with VI [39,40]. Similarly, Audio Virtual Reality (AVR) leverages spatial audio to create immersive auditory environments, translating visual and spatial information into auditory feedback, which is particularly beneficial for students with VI [41,9]. Speech-based interaction technologies, including Text-to-Speech (TTS) and Speech-to-Text (STT), further enhance accessibility by enabling voice commands and auditory feedback in STEAM activities [42]. These advancements are complemented by cutting-edge CV models like CLIP [57], BLIP, YOLO, and Faster R-CNN, which excel in image understanding, object detection, and scene reconstruction, generating real-time auditory descriptions of visual content [43,44]. Together, these technologies—SLAM, AVR, speech-based interactions, and advanced CV models—demonstrate how continuous advancements in computer vision and related fields are driving the development of inclusive educational tools, enabling students with VI to engage more effectively in STEAM activities through multimodal feedback systems.

In summary, while STEAM education is a powerful framework for fostering 21st-century skills, its accessibility for students with VI remains a significant challenge. The visual and spatial nature of many STEAM activities, such as programming, robotics, and laboratory experiments, often excludes students with VI from fully participating. However, advancements in accessible technologies, particularly computer vision (CV), Simultaneous Localization and Mapping (SLAM), Audio Virtual Reality (AVR), and speech-based interaction systems, offer transformative solutions. These technologies can translate visual data into auditory or haptic feedback, enabling students with VI to engage with spatial and visual content in innovative ways. By leveraging these tools, educators can empower students with VI to overcome barriers, develop critical skills, and actively contribute to the STEAM fields, fostering a more equitable and inclusive educational landscape.

## 3. Materials and Methods

This study presents a computer vision-enhanced framework designed to teach students with VI coding and robotics through an accessible, interactive learning environment. The proposed system integrates both overhead and onboard cameras, artificial intelligence (AI)-driven image understanding, depth perception, and speech-based interaction. These components work together to provide real-time awareness of the robot's movement and surroundings, enabling an immersive learning experience.

To facilitate robot programming and navigation, the system incorporates a maze track as a training ground. Mazes have been widely recognised in educational literature as effective tools for developing problem-solving, spatial reasoning, and algorithmic thinking skills [45,46]. By guiding a robot through a structured maze environment, students engage in sequencing, debugging, and logical decision-making, which are fundamental in both programming and robotics education. Additionally, the maze format provides a controlled and adaptable setting, allowing students with VI to receive structured multimodal feedback while improving their understanding of movement and spatial relationships.

The multi-camera setup design, illustrated in Figure 1, consists of a global and local camera system to support students with VI in programming and robotics tasks. The setup includes one or more overhead global cameras, which provide an initial map of the maze environment and track the robot's movement as it executes programmed tasks. Additionally, a local stereo camera mounted on the robot offers a first-person local perspective. This configuration enables CV systems to interpret the surroundings, track navigation, and provide real-time multimodal feedback, such as auditory or haptic cues, enhancing accessibility and engagement in STEAM education.

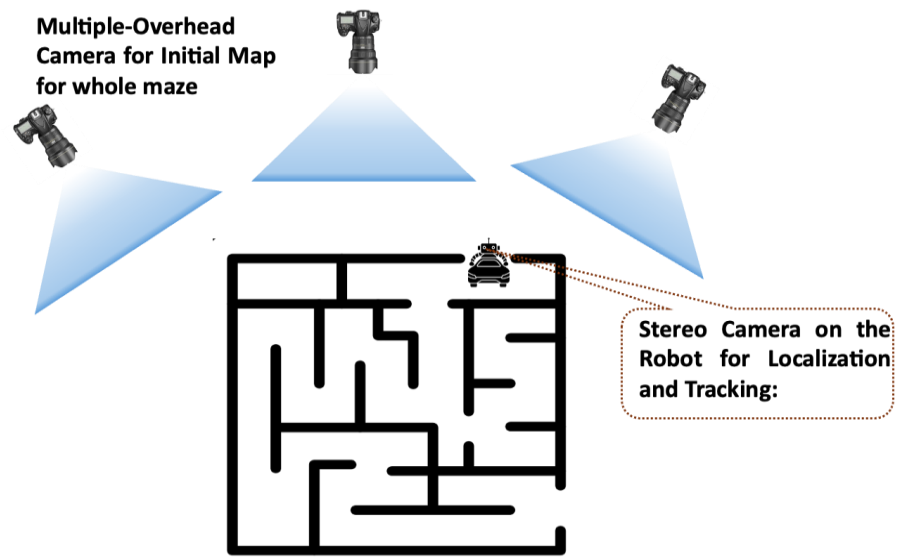

**Figure 1** Multi-camera overhead system capturing the entire maze environment.

The framework comprises three core: global localization via an overhead camera, local navigation using a stereo camera, and AI-based image understanding with speech interaction. The overhead camera captures the maze layout and determines the robot's initial position, essential for accurate task execution and navigation. The captured image undergoes preprocessing to generate a top-down occupancy grid map, aiding localization. This map is then used by computer vision models like CLIP, which convert the environment into textual descriptions. These descriptions are then transformed into spatial audio prompts through an Audio Virtual Reality (AVR) system, providing real-time feedback for students. Figure 2 illustrates the framework's components and data flow for teaching visually impaired students coding and robotics. The system uses an overhead (Global) camera for global localization, a stereo camera for local navigation, and AI models (e.g., CLIP, BLIP) for image understanding. Speech interaction via STT and TTS enables student control and feedback. Real-time obstacle detection, path correction, and safety features ensure adaptive navigation. This multimodal approach creates an accessible, interactive learning experience for visually impaired students.

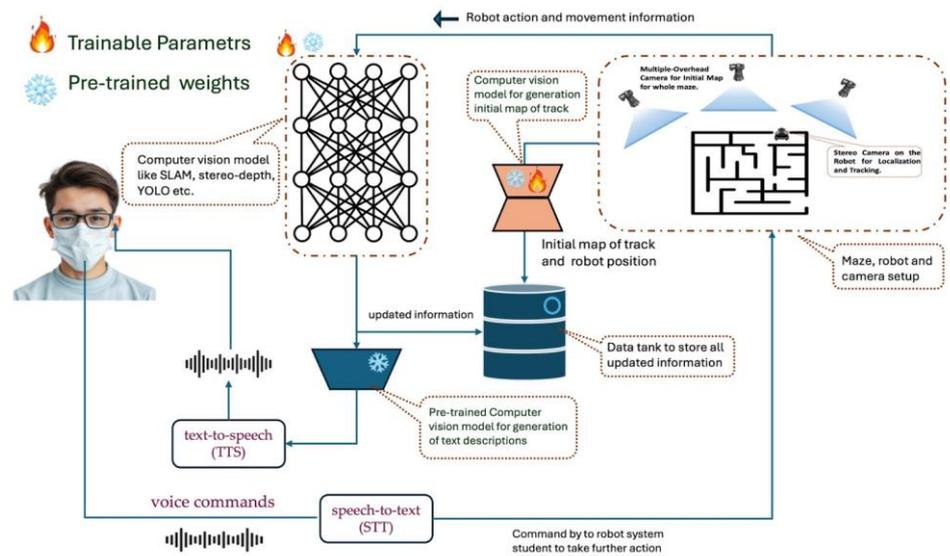

**Figure 2.** System Architecture Overview.

To initiate the robot's movement, students issue voice commands through Speech-to-Text (STT) modules. A stereo camera mounted on the robot captures depth data for obstacle detection and navigation. The real-time data is processed through Simultaneous Localization and Mapping (SLAM) and feature tracking, which improves the robot's path-finding accuracy. The stereo camera also supports object detection and depth estimation using deep learning models such as HitNet [50], RAFT-Stereo [51], Selective Stereo [52], and PatchmatchNet [49]. These processes continuously update the robot's spatial awareness, ensuring effective and adaptive navigation.

Following each predefined movement, computer vision models like BLIP and CLIP [43,47,48] analyse the captured images to generate textual descriptions and calculate the robot's position and orientation relative to its initial position. These descriptions are then converted into natural language audio using Text-to-Speech (TTS) technology. Additionally, the Speech-to-Text (STT) modules allow students to communicate directly with the robot through voice commands, enabling continuous interaction.

In the event of an anomaly, the robot queries the student for guidance. For instance, if an obstacle unexpectedly appears on an initially clear path—such as a block detected on the left while the right and front remain blocked—the robot will prompt, "What should I do? Should I move right?" Once a response is received, the robot carries out the action and updates its position relative to the initial track.

To prevent accidents, such as the robot falling off an edge, a safety threshold is set for the distance gap from obstacles. The stereo camera measures the type and distance of obstacles through stereo depth estimation, ensuring the robot halts before reaching hazardous points. Meanwhile, the global overhead camera continuously monitors the robot's movement, alerting students if the robot deviates from its intended path or initial position.

The data processing pipeline is divided into three key phases. First, in the global localization phase, the overhead camera captures a top-down view of the maze. Image segmentation and edge detection techniques identify track boundaries, and an occupancy grid map is generated to determine the robot's position and orientation. Feature matching algorithms calculate the robot's initial location, and this data is converted into spoken descriptions for student interaction. In the local navigation phase, the stereo camera collects real-time depth data, which is processed using disparity mapping to estimate obstacle distances. Object detection algorithms, such as YOLO [53] or Faster R-CNN [54], identify features in the environment, and their spatial characteristics are conveyed to students through audio descriptions. Robot positioning is continuously updated using Extended Kalman Filter SLAM (EKF-SLAM) for accurate localization. Finally, in the AI-based image understanding and speech interaction phase, images are processed with BLIP and CLIP models to generate textual scene descriptions. These descriptions are converted into speech using TTS, and student voice commands are transcribed using models like Whisper or DeepSpeech. These commands are then translated into Python-based movement instructions, which the robot executes, followed by real-time spoken feedback.

The experimental setup involves a controlled maze environment designed to assess the system's performance. The environment includes a track ranging from 25m² to 50m², with predefined paths and obstacles. It features an overhead camera for global localization and a mobile robot equipped with a stereo camera for local navigation. Participants include a group of blind students learning coding and robotics, along with a control group using traditional tactile-based learning methods. Evaluation metrics include localization accuracy (comparing estimated vs. actual robot position), speech command recognition rate (assessing voice-to-code translation accuracy), task completion time (measuring navigation efficiency), and user experience feedback (evaluating system usability).

Implementation details include both hardware and software components. The system utilizes a 4K wide-angle overhead camera for maze capture and an Intel RealSense D435 stereo camera for depth perception. The robot is a differential-drive mobile unit powered by either a Raspberry Pi or an NVIDIA Jetson-based system. Speech interaction is facilitated through a microphone and speaker. The software stack incorporates ORB-SLAM and EKF-SLAM for localization and mapping, BLIP and CLIP for image understanding, YOLO [53] for object detection, and Whisper, DeepSpeech, and Tacotron 2 for speech processing. The programming interface relies on Python to translate voice commands into executable actions for the robot.

This methodology enables visually impaired students to engage with coding and robotics through multimodal learning, enhancing accessibility and contributing to a more inclusive STEAM education framework.

## 4. Implications and Future Applications

The findings of this study have significant implications for inclusive STEAM education, particularly in enhancing accessibility for students with visual impairments (VI). By integrating computer vision (CV), speech-based interactions, and multimodal feedback, the proposed system provides an inclusive learning environment that allows students with VI to engage with coding and robotics in a meaningful and interactive way. The real-time awareness facilitated by overhead and onboard cameras, combined with AI-driven image understanding, enables students to develop spatial reasoning, logical thinking, and problem-solving skills—key competencies in STEAM education. Furthermore, the introduction of the maze track as a training ground offers structured challenges that reinforce computational thinking and algorithmic problem-solving, aligning with research that highlights mazes as effective educational tools for teaching programming concepts.

Additionally, this framework addresses a key challenge encountered by the authors in previous projects, where students with VI often lost awareness of the robot's location and actions after executing their programmed code. This loss of awareness hindered their ability to debug, modify, and improve their code, limiting their engagement in the iterative problem-solving process essential to programming. By providing continuous real-time feedback through CV and auditory interaction, the proposed system ensures that students remain informed about their robot's movements, allowing them to assess outcomes, identify errors, and make necessary modifications more effectively.

Beyond robotics and programming education, the proposed system has promising applications in Orientation and Mobility (O&M) training, a critical skill set for individuals with VI to navigate their environments safely and independently. By leveraging the system's real-time awareness and AI-driven perception, the framework could be adapted to teach individuals how to use mobility aids effectively, interpret spatial cues, and enhance their navigational skills in controlled and real-world environments. The maze track, originally designed for programming challenges, could serve as a simulated urban setting where users practice route planning, obstacle avoidance, and adaptive decision-making with guided auditory feedback.

Furthermore, this research has the potential to contribute to smart city accessibility solutions by integrating its AI-driven perception and navigation capabilities into urban environments. Smart cities increasingly rely on technology to enhance accessibility, and the system could be extended to support real-time navigation assistance for individuals with VI in public spaces such as transport hubs, pedestrian crossings, and commercial areas. By incorporating computer vision, depth sensing, and auditory feedback, this technology could enhance independent mobility and safety in cities, fostering greater inclusivity and autonomy. Future developments could explore real-world applications such as interactive pedestrian guidance systems, AI-enhanced wayfinding applications, and intelligent transportation integration to improve accessibility in urban environments.

Further research could investigate the integration of haptic feedback and augmented reality (AR) elements to enhance spatial awareness and interaction. Additionally, scaling the system for classroom-wide implementation and evaluating its long-term impact on students' learning outcomes would provide valuable insights into its effectiveness. By advancing accessible educational technologies, this research contributes to fostering an inclusive learning environment where all students, regardless of their abilities, can actively participate in STEAM disciplines and develop essential life skills through immersive, technology-enhanced experiences.

## 5. Limitations and Challenges

Despite the promising potential of the proposed framework, several limitations and challenges must be acknowledged. One of the primary limitations is the technical complexity and setup requirements. The integration of multiple cameras, artificial intelligence-driven perception, and speech-based interaction demands significant computational power and precise hardware calibration. Ensuring seamless real-time processing while maintaining accessibility for students with VI requires optimised system performance, which may pose challenges in resource-limited educational settings.

Another key challenge is dependency on stable and controlled environmental conditions. The effectiveness of the CV system relies on consistent lighting, clear visual markers, and minimised external interference. Variations in classroom lighting or unintentional obstructions may reduce the system's accuracy in detecting the robot's position and surroundings, potentially affecting the learning experience. Future iterations should explore robust algorithms that adapt to dynamic environmental changes to improve reliability.

User adaptation and learning curve also present a challenge, particularly for students and educators unfamiliar with AI-driven assistive technologies. While the system is designed to be accessible, initial training and familiarisation with the interface, auditory feedback, and coding environment may require additional instructional support. The development of user-friendly onboarding materials, including tutorials and guided practice sessions, could help mitigate this challenge.

Additionally, real-time feedback delays and latency issues could impact the effectiveness of the framework, particularly in maintaining continuous awareness of the robot's movements. Any delay in updating the robot's status may lead to confusion and hinder the debugging process, especially for students with VI who rely heavily on auditory cues for spatial awareness. Future improvements should focus on optimising data processing efficiency and exploring edge computing solutions to minimise latency.

From a broader implementation perspective, scalability and cost-effectiveness remain critical factors. Deploying this framework across multiple educational institutions would require affordable hardware options and user-friendly software configurations to ensure widespread accessibility. Further research is needed to explore low-cost alternatives and cloud-based solutions to facilitate broader adoption.

Finally, ethical and privacy considerations must be carefully addressed when implementing CV-enhanced educational tools, particularly in data collection and storage. Ensuring that student interactions and personal data remain secure and anonymised is crucial in maintaining ethical standards. Future iterations of this framework should prioritise robust privacy protocols, user consent mechanisms, and compliance with relevant data protection regulations.

Despite these challenges, the proposed framework represents a significant step towards inclusive robotics education and smart accessibility solutions. Addressing these limitations through iterative development, user feedback, and technological advancements will contribute to a more refined, scalable, and impactful learning experience for students with VI.

## 6. Conclusion

This study presents a comprehensive framework designed to enhance coding and robotics education for students with visual impairments (VI) through a multi-camera, AI-driven interactive system. By integrating both overhead and onboard cameras, artificial intelligence for image understanding, and real-time auditory feedback, the framework provides a unique, accessible learning environment that promotes autonomy and engagement. The system's design, with its real-time awareness of the robot's movements and surroundings, addresses key challenges faced by students with VI, particularly in the context of maintaining spatial awareness and enabling effective debugging.

The proposed framework offers significant educational potential, especially in providing hands-on learning opportunities for students with VI to develop essential skills in coding, robotics, and problem-solving. Furthermore, it opens the door to future applications in orientation and mobility (O&M) training, allowing students to learn how to navigate and use mobility aids in real-world environments. The potential for this framework to contribute to the accessibility of smart cities further underscores its relevance in the evolving landscape of assistive technologies.

However, several limitations and challenges remain, including technical complexity, environmental factors, user adaptation, and scalability. Despite these challenges, the framework offers a promising pathway for future development, with potential improvements in real-time feedback, system reliability, and cost-effective implementation. By addressing these challenges, future iterations of the framework could provide an even more impactful and widely accessible tool for enhancing educational outcomes for students with VI.

In conclusion, this framework represents a significant step toward more inclusive educational practices, empowering students with visual impairments to engage with coding, robotics, and other STEM fields in a manner that fosters independence, creativity, and problem-solving. With ongoing development and iteration, this framework has the potential to become an integral tool in accessible education and broader societal applications, such as in smart cities and mobility aids.

**Author Contributions:** Conceptualization, M.H. and MD. K.; methodology, MD. K.; software, MD. K.; validation, M.H. and MD. K. and P.T.;; writing—original draft preparation, M.H. and MD. K.; writing—review and editing, M.H. and MD. K.; visualization, MD. K.; project administration, M.H. All authors have read and agreed to the published version of the manuscript.

**Funding:** This research received no external funding.

**Institutional Review Board Statement:** Not applicable.

**Informed Consent Statement:** Not applicable.
**Data Availability Statement:** Not applicable.

**Acknowledgements:** Not applicable.

**Conflicts of Interest:** The authors declare no conflicts of interest.

## Abbreviations

The following abbreviations are used in this manuscript:

STEAM    Science, Technology, Engineering, Arts, and Mathematics

VI    Visual Impairment

CLIP    Contrastive Language-Image Pre-training

AVR    Audio Virtual Reality

SLAM            Simultaneous Localization and Mapping

# References

1. Sanders, M. (2009). STEM, STEM education, STEMmania. The Technology Teacher, December/January.
2. Perignat, E., & Katz-Buonincontro, J. (2019). STEAM in practice and research: An integrative literature review. Thinking Skills and Creativity, 31, 31–43. https://doi.org/10.1016/j.tsc.2018.10.002
3. Burgstahler, S. (2011). Universal Design: Implications for Computing Education. ACM Transactions on Computing Education, 11(3), 1–17. https://doi.org/10.1145/2037276.2037283
4. Hamash, M., & Mohamed, H. (2021). BASAER Team: The First Arabic Robot Team for Building the Capacities of Visually Impaired Students to Build and Program Robots. International Journal of Emerging Technologies in Learning (iJET), 16(24), 91–107. https://doi.org/10.3991/ijet.v16i24.27465
5. Supalo, C. A., Humphrey, J. R., Mallouk, T. E., David Wohlers, H., & Carlsen, W. S. (2016). Examining the use of adaptive technologies to increase the hands-on participation of students with blindness or low vision in secondary-school chemistry and physics. Chemistry Education Research and Practice, 17(4), 1174–1189. https://doi.org/10.1039/C6RP00141F
6. Eguchi, A. (2014). Educational Robotics for Promoting 21st Century Skills. Journal of Automation, Mobile Robotics and Intelligent Systems, 5–11. https://doi.org/10.14313/JAMRIS_1-2014/1
7. Atmatzidou, S., & Demetriadis, S. (2016). Advancing students' computational thinking skills through educational robotics: A study on age and gender relevant differences. Robotics and Autonomous Systems, 75, 661–670. https://doi.org/10.1016/j.robot.2015.10.008
8. Kumar, K. L., & Wideman, M. (2014). Accessible by design: Applying UDL principles in a first year undergraduate course. Canadian Journal of Higher Education, 44(1), 125–147. https://doi.org/10.47678/cjhe.v44i1.183704
9. Hamash, M., Ghreir, H., & Tiernan, P. (2024). Breaking through Barriers: A Systematic Review of Extended Reality in Education for the Visually Impaired. Education Sciences, 14(4), 365. https://doi.org/10.3390/educsci14040365
10. Ludi, S. L., Ellis, L., & Jordan, S. (2014). An accessible robotics programming environment for visually impaired users. Proceedings of the 16th International ACM SIGACCESS Conference on Computers & Accessibility - ASSETS '14, 237–238. https://doi.org/10.1145/2661334.2661385
11. Milne, L. R., & Ladner, R. E. (2018). Blocks4All: Overcoming Accessibility Barriers to Blocks Programming for Children with Visual Impairments. Proceedings of the 2018 CHI Conference on Human Factors in Computing Systems, 1–10. https://doi.org/10.1145/3173574.3173643
12. Basham, J. D., & Marino, M. T. (2013). Understanding STEM Education and Supporting Students through Universal Design for Learning. TEACHING Exceptional Children, 45(4), 8–15. https://doi.org/10.1177/004005991304500401
13. Israel, M., Wherfel, Q. M., Pearson, J., Shehab, S., & Tapia, T. (2015). Empowering K–12 Students With Disabilities to Learn Computational Thinking and Computer Programming. TEACHING Exceptional Children, 48(1), 45–53. https://doi.org/10.1177/0040059915594790
14. Metatla, O., Serrano, M., Jouffrais, C., Thieme, A., Kane, S., Branham, S., Brulé, É., & Bennett, C. L. (2018). Inclusive Education Technologies: Emerging Opportunities for People with Visual Impairments. Extended Abstracts of the 2018 CHI Conference on Human Factors in Computing Systems, 1–8. https://doi.org/10.1145/3170427.3170633
15. Istenic Starcic, A., & Bagon, S. (2014). ICT -supported learning for inclusion of people with special needs: Review of seven educational technology journals, 1970–2011. British Journal of Educational Technology, 45(2), 202–230. https://doi.org/10.1111/bjet.12086
16. Khan, M. R., Negi, A., Kulkarni, A., Phutke, S. S., Vipparthi, S. K., & Murala, S. (2024). Phaseformer: Phase-based Attention Mechanism for Underwater Image Restoration and Beyond (Version 1). arXiv. https://doi.org/10.48550/ARXIV.2412.01456
17. Sivan, S., & Darsan, G. (2016). Computer Vision based Assistive Technology for Blind and Visually Impaired People. Proceedings of the 7th International Conference on Computing Communication and Networking Technologies, 1–8. https://doi.org/10.1145/2967878.2967923
18. Terven, J. R., Salas, J., & Raducanu, B. (2014). New Opportunities for Computer Vision-Based Assistive Technology Systems for the Visually Impaired. Computer, 47(4), 52–58. https://doi.org/10.1109/MC.2013.265
19. Hamash, M., & AbuSuliman, N. (2021). Advanced practical Guide for Building the Capacities of Visually Impaired Students to Build and Program Robots (Vol. 01). Amazon.com.ca. https://www.amazon.ca/Advanced-practical-Building-Capacities-Visually-ebook/dp/B09HJWJ5MH

20. White, J. L., & Massiha, G. H. (2015). Strategies to Increase Representation of Students with Disabilities in Science, Technology, Engineering and Mathematics (STEM). International Journal of Evaluation and Research in Education (IJERE), 4(3), 89. https://doi.org/10.11591/ijere.v4i3.4497
21. Hacıoğlu, Y., & Suiçmez, E. (2022). STEAM education in preschool education: We design our school for our visually impaired friend. Science Activities, 59(2), 55–67. https://doi.org/10.1080/00368121.2022.2056111
22. Baker, C. M., Milne, L. R., & Ladner, R. E. (2015). StructJumper: A Tool to Help Blind Programmers Navigate and Understand the Structure of Code. Proceedings of the 33rd Annual ACM Conference on Human Factors in Computing Systems, 3043–3052. https://doi.org/10.1145/2702123.2702589
23. Howard, A. M., Chung Hyuk Park, & Remy, S. (2012). Using Haptic and Auditory Interaction Tools to Engage Students with Visual Impairments in Robot Programming Activities. IEEE Transactions on Learning Technologies, 5(1), 87–95. https://doi.org/10.1109/TLT.2011.28
24. Stefik, A. M., Hundhausen, C., & Smith, D. (2011). On the design of an educational infrastructure for the blind and visually impaired in computer science. Proceedings of the 42nd ACM Technical Symposium on Computer Science Education, 571–576. https://doi.org/10.1145/1953163.1953323
25. Ludi, S., & Jordan, S. (2015). A JBrick: Accessible Robotics Programming for Visually Impaired Users. In M. Antona & C. Stephanidis (Eds.), Universal Access in Human-Computer Interaction. Access to Learning, Health and Well-Being (Vol. 9177, pp. 157–168). Springer International Publishing. https://doi.org/10.1007/978-3-319-20684-4_16
26. Brulé, E., Tomlinson, B. J., Metatla, O., Jouffrais, C., & Serrano, M. (2020). Review of Quantitative Empirical Evaluations of Technology for People with Visual Impairments. Proceedings of the 2020 CHI Conference on Human Factors in Computing Systems, 1–14. https://doi.org/10.1145/3313831.3376749
27. Szeliski, R. (2022). Computer vision: Algorithms and applications (Second edition). Springer.
28. Ponce, J. (Ed.). (2012). Computer Vision: A Modern Approach: International Edition (2. Auflage). Pearson Education, Limited.
29. Khan, M. R., Kulkarni, A., Phutke, S. S., & Murala, S. (2023). Underwater Image Enhancement with Phase Transfer and Attention. 2023 International Joint Conference on Neural Networks (IJCNN), 1–8. https://doi.org/10.1109/IJCNN54540.2023.10191620
30. Brown, C. M., & Ballard, D. H. (1982). Computer Vision. Prentice Hall.
31. LeCun, Y., Bengio, Y., & Hinton, G. (2015). Deep Learning. Nature, 521(7553), 436–444. https://doi.org/10.1038/nature14539
32. Wang, H.-C., Katzschmann, R. K., Teng, S., Araki, B., Giarre, L., & Rus, D. (2017). Enabling independent navigation for visually impaired people through a wearable vision-based feedback system. 2017 IEEE International Conference on Robotics and Automation (ICRA), 6533–6540. https://doi.org/10.1109/ICRA.2017.7989772
33. Tapu, R., Mocanu, B., & Zaharia, T. (2020). Wearable assistive devices for visually impaired: A state of the art survey. Pattern Recognition Letters, 137, 37–52. https://doi.org/10.1016/j.patrec.2018.10.031
34. Brock, M., & Kristensson, P. O. (2013). Supporting blind navigation using depth sensing and sonification. Proceedings of the 2013 ACM Conference on Pervasive and Ubiquitous Computing Adjunct Publication, 255–258. https://doi.org/10.1145/2494091.2494173
35. Shoval, S., Ulrich, I., & Borenstein, J. (2003). Robotics-based obstacle-avoidance systems for the blind and visually impaired—Navbelt and the guidecane. IEEE Robotics & Automation Magazine, 10(1), 9–20. https://doi.org/10.1109/MRA.2003.1191706
36. Sánchez, J., & Aguayo, F. (2005). Blind learners programming through audio. CHI '05 Extended Abstracts on Human Factors in Computing Systems, 1769–1772. https://doi.org/10.1145/1056808.1057018
37. Landau, S. (2021). Development of a Talking Tactile Tablet. Information Technology and Disabilities Journal, 7(2). http://itd.athenpro.org/volume7/number2/tablet.html
38. Sharif, A., Wang, O. H., Muongchan, A. T., Reinecke, K., & Wobbrock, J. O. (2022). VoxLens: Making Online Data Visualizations Accessible with an Interactive JavaScript Plug-In. CHI Conference on Human Factors in Computing Systems, 1–19. https://doi.org/10.1145/3491102.3517431
39. Durrant-Whyte, H., & Bailey, T. (2006). Simultaneous localization and mapping: Part I. IEEE Robotics & Automation Magazine, 13(2), 99–110. https://doi.org/10.1109/MRA.2006.1638022
40. Cadena, C., Carlone, L., Carrillo, H., Latif, Y., Scaramuzza, D., Neira, J., Reid, I., & Leonard, J. J. (2016). Past, Present, and Future of Simultaneous Localization and Mapping: Toward the Robust-Perception Age. IEEE Transactions on Robotics, 32(6), 1309–1332. https://doi.org/10.1109/TRO.2016.2624754

41. Rajguru, C., Brianza, G., & Memoli, G. (2022). Sound localization in web-based 3D environments. Scientific Reports, 12(1), 12107. https://doi.org/10.1038/s41598-022-15931-y
42. Sánchez, J., Sáenz, M., Pascual-Leone, A., & Merabet, L. (2010). Navigation for the blind through audio-based virtual environments. CHI '10 Extended Abstracts on Human Factors in Computing Systems, 3409–3414. https://doi.org/10.1145/1753846.1753993
43. Radford, A., Kim, J. W., Hallacy, C., Ramesh, A., Goh, G., Agarwal, S., Sastry, G., Askell, A., Mishkin, P., Clark, J., Krueger, G., & Sutskever, I. (2021). Learning Transferable Visual Models From Natural Language Supervision (Version 1). arXiv. https://doi.org/10.48550/ARXIV.2103.00020
44. Redmon, J., Divvala, S., Girshick, R., & Farhadi, A. (2015). You Only Look Once: Unified, Real-Time Object Detection (Version 5). arXiv. https://doi.org/10.48550/ARXIV.1506.02640
45. Ludi, S. A., & Reichlmayr, T. (2008). Developing inclusive outreach activities for students with visual impairments. ACM SIGCSE Bulletin, 40(1), 439–443. https://doi.org/10.1145/1352322.1352285
46. Buck, J. (2015). Mazes for programmers: Code your own twisty little passages (J. Carter, Ed.). The Pragmatic Bookshelf.
47. Rao, R., Meier, J., Sercu, T., Ovchinnikov, S., & Rives, A. (2020). Transformer protein language models are unsupervised structure learners. https://doi.org/10.1101/2020.12.15.422761
48. Singh, A., Hu, R., Goswami, V., Couairon, G., Galuba, W., Rohrbach, M., & Kiela, D. (2021). FLAVA: A Foundational Language And Vision Alignment Model (Version 3). arXiv. https://doi.org/10.48550/ARXIV.2112.04482
49. Wang, F., Galliani, S., Vogel, C., Speciale, P., & Pollefeys, M. (2021). Patchmatchnet: Learned multi-view patchmatch stereo. In Proceedings of the IEEE/CVF conference on computer vision and pattern recognition (pp. 14194-14203).
50. Tankovich, V., Hane, C., Zhang, Y., Kowdle, A., Fanello, S., & Bouaziz, S. (2021). Hitnet: Hierarchical iterative tile refinement network for real-time stereo matching. In Proceedings of the IEEE/CVF Conference on Computer Vision and Pattern Recognition (pp. 14362-14372).
51. Lipson, L., Teed, Z., & Deng, J. (2021, December). Raft-stereo: Multilevel recurrent field transforms for stereo matching. In 2021 International Conference on 3D Vision (3DV) (pp. 218-227). IEEE.
52. Wang, X., Xu, G., Jia, H., & Yang, X. (2024). Selective-stereo: Adaptive frequency information selection for stereo matching. In Proceedings of the IEEE/CVF Conference on Computer Vision and Pattern Recognition (pp. 19701-19710).
53. Hussain, M. (2023). YOLO-v1 to YOLO-v8, the rise of YOLO and its complementary nature toward digital manufacturing and industrial defect detection. Machines, 11(7), 677
54. Ren, S., He, K., Girshick, R., & Sun, J. (2016). Faster R-CNN: Towards real-time object detection with region proposal networks. IEEE transactions on pattern analysis and machine intelligence, 39(6), 1137-1149.
55. MAILKA, H., Abouzahir, M., & Ramzi, M. (2024). An efficient end-to-end EKF-SLAM architecture based on LiDAR, GNSS, and IMU data sensor fusion for autonomous ground vehicles. Multimedia Tools and Applications, 83(18), 56183-56206.
56. Ni, J., Wang, X., Gong, T., & Xie, Y. (2022). An improved adaptive ORB-SLAM method for monocular vision robot under dynamic environments. International Journal of Machine Learning and Cybernetics, 13(12), 3821-3836.
57. Li, Y., Liang, F., Zhao, L., Cui, Y., Ouyang, W., Shao, J., ... & Yan, J. (2021). Supervision exists everywhere: A data efficient contrastive language-image pre-training paradigm. arXiv preprint arXiv:2110.05208.